\documentclass[mathleft
]{an}
\usepackage{graphicx}
\usepackage{times}
\usepackage{multirow}

\def\xmm{{\it XMM--Newton}}

\def\nustar{{\it NuSTAR}}
\def\athena{{\it Athena}}

\newcommand{\xr}{X--ray}
\newcommand{\kms}{km\,s$^{-1}$}
\newcommand{\ie}{i.e.}
\newcommand{\eg}{e.g.}

\newcommand{\kev}{keV}

\newcommand{\etal}{et al.}

\newcommand{\swi}{SWIFT\,J2127.4+5654}

\usepackage{natbib}
\bibpunct{(}{)}{;}{a}{}{,}
\begin{document}

\Pagespan{1}{6}
\Yearpublication{2015}%
\Yearsubmission{2015}%
\Month{6}%
\Volume{--}%
\Issue{--}%
\DOI{--}%

\title{Eclipsing the innermost accretion disc regions in AGN}

\author{M.~Sanfrutos\inst{1}\fnmsep\thanks{Corresponding author:
  \email{sanfrutoscm@cab.inta--csic.es}\newline}
\and  G.~Miniutti\inst{1}
\and  M.~Dov\v{c}iak\inst{2}
\and  B.~Ag\'is--Gonz\'alez\inst{1}
}
\titlerunning{Eclipsing the innermost \xr\ emitting regions in AGN}
\authorrunning{M.~Sanfrutos \etal}
\institute{Centro de Astrobiolog\'{i}a (CSIC--INTA), Dep. de Astrof\'{i}sica; 
ESAC, PO Box 78, Villanueva de la Ca\~nada, E-28691 Madrid, Spain
\and
Astronomical Institute, Academy of Sciences, Bo\v{c}n\'i II, CZ-14131 Prague, Czech Republic}

\received{1 Sep 2015}
\accepted{22 Sep 2015}
\publonline{later}


\abstract{
Variable \xr\ absorption has been observed in active galactic nuclei (AGN) on several time scales. Observations allow us to identify the absorber with clouds associated either with the clumpy torus (parsec scales, long timescales) or with the broad line region (BLR) (short timescales). In the latter, the cloud size has been estimated to be of the order of few gravitational radii from the observed absorption variability. Such small cloud sizes are comparable to the \xr\ emitting regions so that a detailed modeling of occultation events in AGN has the potential of enabling us to infer accurately the geometry of the system. We have developed a relativistic \xr\ spectral model for occultation events and we present here theoretical predictions on the different observables that can be inferred by studying \xr\ eclipses in simulated \xmm\ data. These include the size of the \xr\ emitting regions as well as more fundamental parameters such as the black hole spin and the system inclination. We find that absorption varies as a function of the energy range and that its maximum takes place when the approaching part of the accretion disc is covered. Therefore we study the hard--to--soft (H\,/\,S) ratio light curves produced during an eclipse and use them to characterise the properties of the inner accretion disc in a new model--independent way. 
}

\maketitle

\section{Introduction}
\label{sec:intro}

Several spectral features shuch as Doppler boosting, gravitational redshift and light bending are accounted for in the \xr\ emission reprocessed by the innermost regions of the accretion disc, which are in a strong gravity regime due to its proximity to the supermassive black hole (SMBH). In this context, both the black hole spin and the disc properties shape the K$\alpha$ emission line profile \citep[see e.g.][]{Reynolds03}. 
However, there is some controversy on this interpretation: the partial covering scenario supported by \cite{Turner07} claims that such spectral features can be explained by calling upon a variable wind composed by several ionised clouds that partially cover the \xr\ continuum source. 
Within this paper we argue that the ultimate test to prove the detectability of general relativistic effects under extreme gravity conditions could be furnished by \xr\ spectroscopy of AGN, which would definitely rule out the partial covering scenario. 

On the side of special relativistic effects, Doppler boosting would have a clear and distinct effect on the reflected component: the emission from the parts of the disc approaching towards us would be enhanced while we would detect diminished flux from the receding parts. Due to the fact that the central engine is unresolved, 
the only way to distinguish the reflection from the various parts of the disc is by means of the obscuration of the different emitting regions by structures in our line of sight (LOS). 
Occultation of the \xr\ source by optically thick matter was first proposed by \cite{mckernan98}. 
A method to investigate relativistic effects during \xr\ eclipses in AGN spectra was developed by \cite{risaliti2011relat}, consisting on measuring the emission from half of the disc while the other half is covered, and viceversa. Relativistic effects arise when the emission of each half are different, hinting an asymmetric emission from the disc due to Doppler boosting. 

Occultation events are rather common in AGN. 
These objects show \xr\ spectral variability on a variety of time--scales, 
independently of their luminosity or morphology, 
\citep[see e.g. ][]{marshall81}. 
Specifically, Narrow Line Seyfert\,1 (NLS1)  
galaxies systematically show a higher amplitude variability 
\citep{leighly99_nls1}. 
Also, variability can occur at time--scales as short as a few hours in all types of AGN, as shown by 
\cite{elvis2004} in NGC\,4388 (type\,2), 
\cite{bianchi2009} in NGC\,7582 (type\,2), 
\cite{Risaliti09} and \cite{maiolino10} in NGC\,1365 (Seyfert\,1.8), 
\cite{puccetti2007} in NGC\,4151 (Seyfert\,1.5), 
\cite{Risaliti11_mrk766} in Mrk\,766 (NLS1) and 
\cite{sanfrutos2013} in \swi\ (NLS1). 
The high variability characterised in the references above can be explained in the context of rapid column density changes, which are the result of the interception of material located in the BLR of the system to our LOS
\citep[see e.g.][]{miniutti2014}. 
Hence, AGN are perfect laboratories to check general relativity (GR) effects by using eclipses to probe the innermost regions of the accretion disc. 

%
\section{The absorption model}
\label{sec:absorption_model}

The geometry that we assume for the system consists of a SMBH, characterised by its mass ($M_{\rm BH}$) and spin parameter (${\rm a^*}$), around which an optically thick but physically thin ionised accretion disc extends from the innermost stable circular orbit (ISCO) to $1000\,r_{\rm g}$. 
About the geometry of the system formed by the disc and the \xr\ source (or the so--called corona), several configurations have been provided in the literature: 
the hot and radiatively compact spherical corona with radius between $3$ and $10\,r_{\rm g}$ proposed by \cite{Fabian15}; 
the slab \citep{rozanska15} 
or the patchy \citep{stern95} geometries; 
and the jet base interpretation \citep{Wilkins15b}, 
just to mention some of them. Our goal in this work is only to perform a first exploration of the inner accretion disc obscuration, therefore we do not choose any particular geometry for the corona yet, leaving that point to a more extensive work (Sanfrutos \etal\ in preparation). However, the continuum \xr\ source has to be located in the vicinity of the SMBH in order to properly illuminate the innermost regions of the accretion disc; 
\ie \ in our approximation we assume that both the primary continuum and the reflection component arise together from every point on the disc. 
 
The eclipsing cloud is assumed to be co--rotating with the disc at a velocity of 3000\,\kms , and therefore to be located much further away ($10^4\,r_{\rm g}$), in the BLR of the system. 
Since these kind of obscuration events are more likely to be detected in unobscured sources, we disregard the torus of the Unification models within the framework of this paper. 

The model we are using consists of 
a power law plus an \xr\ reflection component arising from the accretion disc, which is represented by the {\small XILLVER} code in {\small{XSPEC}} \citep{Garcia13}. 
These two components are multiplied by the {\small KYNCONV} relativistic convolution model from \cite{dovciak2004}, that produces accretion discs spectra in the strong gravity regime, allowing to obscure part of the emission with a circular cloud whose size and position can be determined. 
We model the accretion disc to cover at least one half of the sky as seen from the central engine, with a typical ionisation of $\xi^{\rm (disc)} = 100$\,erg\,cm\,s$^{-1}$ and solar abundances. The energy cutoff is frozen to 300\,\kev . 
The main parameters that can be tuned in the {\small KYNCONV} relativistic convolution model are: 
(i) the inclination $\theta$ of the system with respect to our LOS ($0\degr$: face--on, $89\degr$: edge--on); 
(ii) the black hole spin ${\rm a}^*$ ($0$: not rotating, $0.998$: maximally rotating); 
(iii) the emissivity index ${\rm q}$, giving account of how the disc is illuminated (low: uniform, high: centrally concentrated); 
(iv) the cloud position set by the impact parameters $\alpha$ and $\beta$ in units of $r_{\rm g}$ with horizontal position being positive towards the approaching side of the accretion disc and negative towards the receding part, and with zero vertical position of the cloud (\ie\ $\beta=0$); 
and (v) the cloud size in terms of its radius $r$, also in $r_{\rm g}$ units. 
The intrinsical physical properties of the obscuring cloud such as its column density $N_{\rm H}^{\rm (cl.)}$ and ionisation log\,$\xi^{\rm (cl.)}$ are set by muliplying the components defined above by the ionised absorption code {\small{ZXIPCF}} \citep{reeves_zxipcf2008}. 
We take into consideration the Galactic absorption too, by means of a neutral absorption component. 
We choose an intermediate arbitrary value of $N_{\rm H}^{\rm (Gal.)} = 3 \times 10^{21}\,{\rm cm}^{-2}$. 

The procedure we perform is the following: for every position of the cloud we compute the flux in the $0.7-10$\,\kev\ energy band, then we compute the intrinsic flux in the same band by removing the cloud, and finally we derive the covering fraction (CF) at every position of the cloud all along the eclipse. Any asymmetry in these CF--light curves would be indicative of asymmetries in the emission from the disc, and therefore it would evidence relativistic effects. 

%
\subsection{The Compton--thick neutral absorber}
\label{subsec:cthick}

\begin{figure*}
\begin{center}
\includegraphics[width=\textwidth]{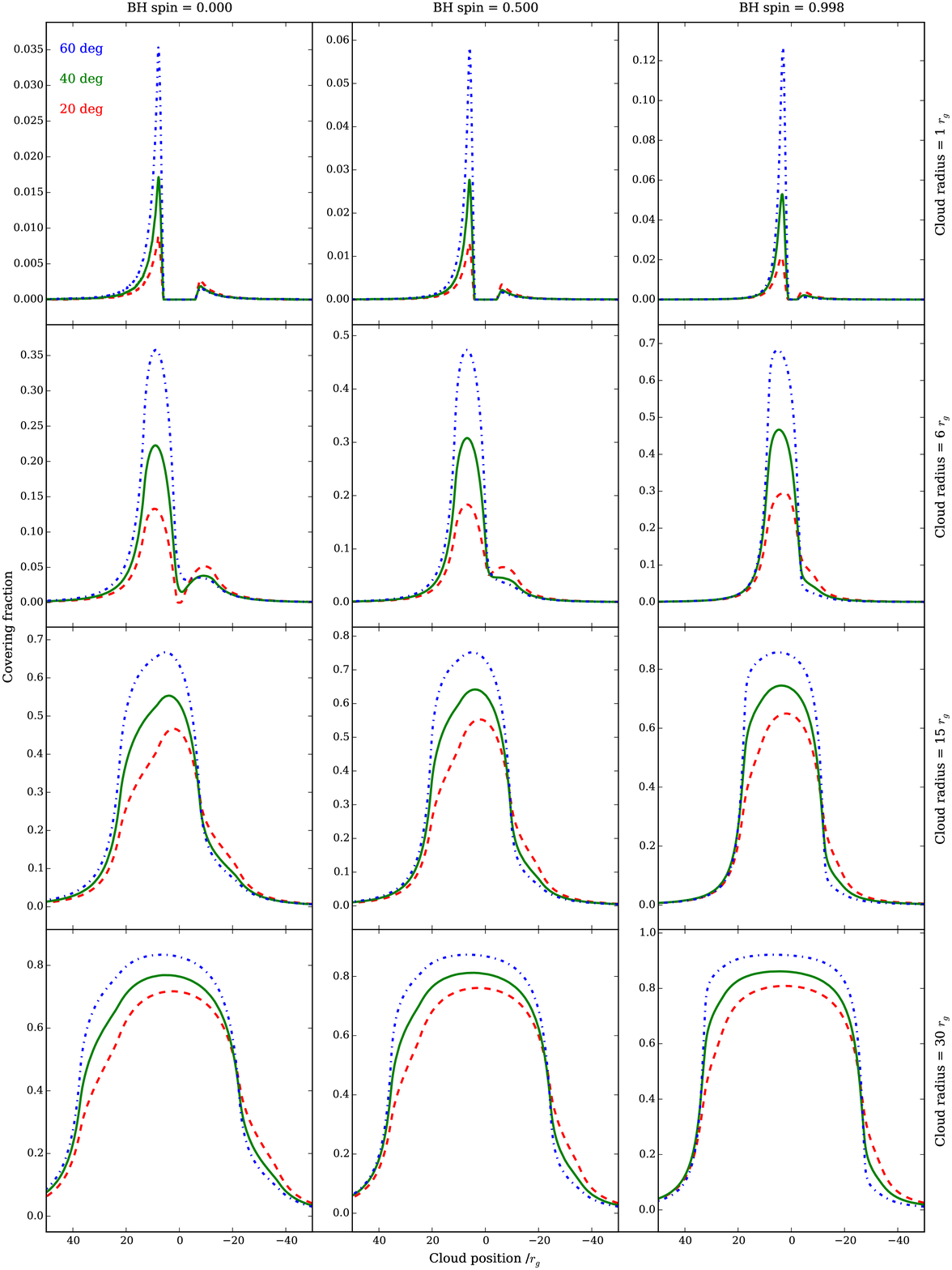}
\caption{\label{fig:cf_profiles}A bunch of 36 CF--profiles produced by a neutral cloud of column density $N_{\rm H} = 10^{24}$\,cm$^{-2}$ obscuring a disc illuminated with an emissivity index of ${\rm q} = 3$. The black hole spin ranges from 0 on the left column to 0.998 on the right. The cloud size ranges from $1\,r_{\rm g}$ on the upper row up to $30\,r_{\rm g}$ on the lower. We observe the system at $20 \degr$ (dashed line, red online), $40 \degr$ (solid line, green online) and $60 \degr$ (dash--dotted line, blue online) with respect to the disc's normal. See Section\,\ref{subsec:cthick} for a detailed explanation. }
\end{center}
\end{figure*}

Both low column densities and large ionisations are associated with more transparent clouds, and therefore with fainter CFs. Since our goal is to study the shape of the CF--profiles in order to look for asymmetries pointing to relativistic effects, we decided to explore the parameters space of an absorber able to produce noticeable coverings of the flux emitted by the accretion disc. Hence, the parameters of the absorber that we consider are those of an unphysical opaque neutral cloud, \ie\ $N_{\rm H}^{\rm (cl.)} = 10^{24}$\,cm$^{-2}$ and log\,$\xi^{\rm (cl.)} = -3$. The study of realistic clouds (Compton--thin and Compton--thick up to $5\times 10^{24}$\,cm$^{-2}$, and ionisations of up to log\,$\xi^{\rm (cl.)} = 6$) is beyond the purpose of this paper and will be detailed elsewhere (Sanfrutos \etal\ in preparation). 

The illumination profiles of the disc play an important role too: the steeper the emissivity index is, the more centrally concentrated is the corona and the flux is reflected by material within regions closer to the central engine. Therefore, the higher is the emissivity index parameter ${\rm q}$ the deeper is the eclipse and the larger the CF. Below we present results from some CF--profiles computed with an intermediate emissivity index of ${\rm q} = 3$. The CF--profiles are compiled in Fig.\,\ref{fig:cf_profiles}. 

The first parameter about which we can establish some conclusions is the inclination of the system, in terms of the angle between the normal to the disc and our LOS. When we observe the system face--on (\ie\ $0 \degr$), Doppler boosting is undetectable from our point of view: there are no approaching nor receding regions in the disc, since its orbital plane is perpendicular to our LOS. Therefore the CF--profile observed would be symmetric with respect to the position of the black hole. The larger the inclination is, the greater is the Doppler boosting effect, so that the more noticeable is the asymmetry of the CF--profiles, as can be seen in every panel of Fig.\,\ref{fig:cf_profiles}, independently of every other parameter. The CF--profiles during an eclipse in a system at an inclination of $20 \degr$ (near to pole--on) are shown in dashed red lines; at a greater inclination of $40 \degr$ in solid green lines, showing a greater asymmetry; and finally at an inclination of $60 \degr$ in dash--dotted blue lines, when the asymmetry is the largest. We do not show inclinations lower than $20 \degr$ because one would not expect to find BLR clouds at such high latitudes. We do not show inclinations larger than $60 \degr$ either because the torus would make impossible any observation. 

The space of parameters was explored for three characteristic values of the spin. On the left column of Fig.\,\ref{fig:cf_profiles} we show the CF--profiles of a Schwarzschild black hole (ISCO = 6\,$r_{\rm g}$). On the column in the middle we show the same for an intermediate Kerr black hole (ISCO = 4.23\,$r_{\rm g}$). On the right column we show the CFs with a maximally rotating Kerr black hole (ISCO = 1.25\,$r_{\rm g}$). In the cases where the cloud size is comparable to the ISCO (upper row in Fig.\,\ref{fig:cf_profiles}, with a cloud radius of 1\,$r_{\rm g}$, and also in the left and even the middle panels in the second row, with a radius of 6\,$r_{\rm g}$), we could measure the ISCO with great precision in a completely new way when the size of the cloud is known, just by measuring the distance between the two maxima in any of those panels. When larger clouds are involved in the eclipse, general relativistic effects are still detectable, as highlights the asymmetry of the characteristic top hat--shaped CF--profiles shown on any of the panels in the third ($r = 15\,r_{\rm g}$) and fourth ($r = 30\,r_{\rm g}$) rows in Fig.\,\ref{fig:cf_profiles}. 

It is important to mention that some of the maxima in the CF--profiles shown in Fig.\,\ref{fig:cf_profiles} are beyond the limits of detectability of available instruments (mainly the upper row cases, when the cloud radius is too small). However, most of them still represent interesting cases that we could discern by using current technology. 

%
\section{Spectral analysis}
\label{sec:spectral_analysis}

We focus now on the case of an intermediate Kerr black hole seen at an inclination of $40 \degr$ and obscured by two different clouds: one ($r = 6\,r_{\rm g}$) slightly larger than the ISCO and one slightly smaller ($r = 3\,r_{\rm g}$). We choose a relatively steep emissivity index ${\rm q} = 6$ in order to get a more centrally concentrated emission from the accretion disc, which will allow the occultation event to reach noticeable CFs with small--sized clouds. 
Such steep emissivities or even larger were reported in some of the AGN, \eg\ 1H\,0707-495 \citep{Fabian09} or IRAS13224-3809 \citep{Ponti10}. 

In order to check if the existence of such a cloud is physically possible attending to the physical constraints of the \xr\ source and absorber sizes and the distance between them, we assume that the central SMBH mass is $10^7\,M_{\odot}$, hence its gravitational radius is $r_{\rm g} = 10^{12}$\,cm. With a cloud of column density $N_{\rm H} = 10^{24}$\,cm$^{-2}$, this implies densities of $8 \times 10^{10}$\,cm$^{-3}$ and $1.7 \times 10^{11}$\,cm$^{-3}$ respectively, which are typical values for the clouds in the BLR. Assuming an orbital velocity of $3000$\,\kms , as already mentioned in Section\,\ref{sec:absorption_model}, the cloud would be located at a distance of $10^4\,r_{\rm g} = 10^{16}\,$cm from the SMBH, and would travel the region of our study ($\sim 40\,r_{\rm g}$) in $\sim 133$\,ks, at a velocity of $0.3\,r_{\rm g}$\,ks$^{-1}$. 

\subsection{Simulated spectra}
\label{subsec:simulated_spectra}

\begin{figure}
\begin{center}
\includegraphics[width=0.48\textwidth]{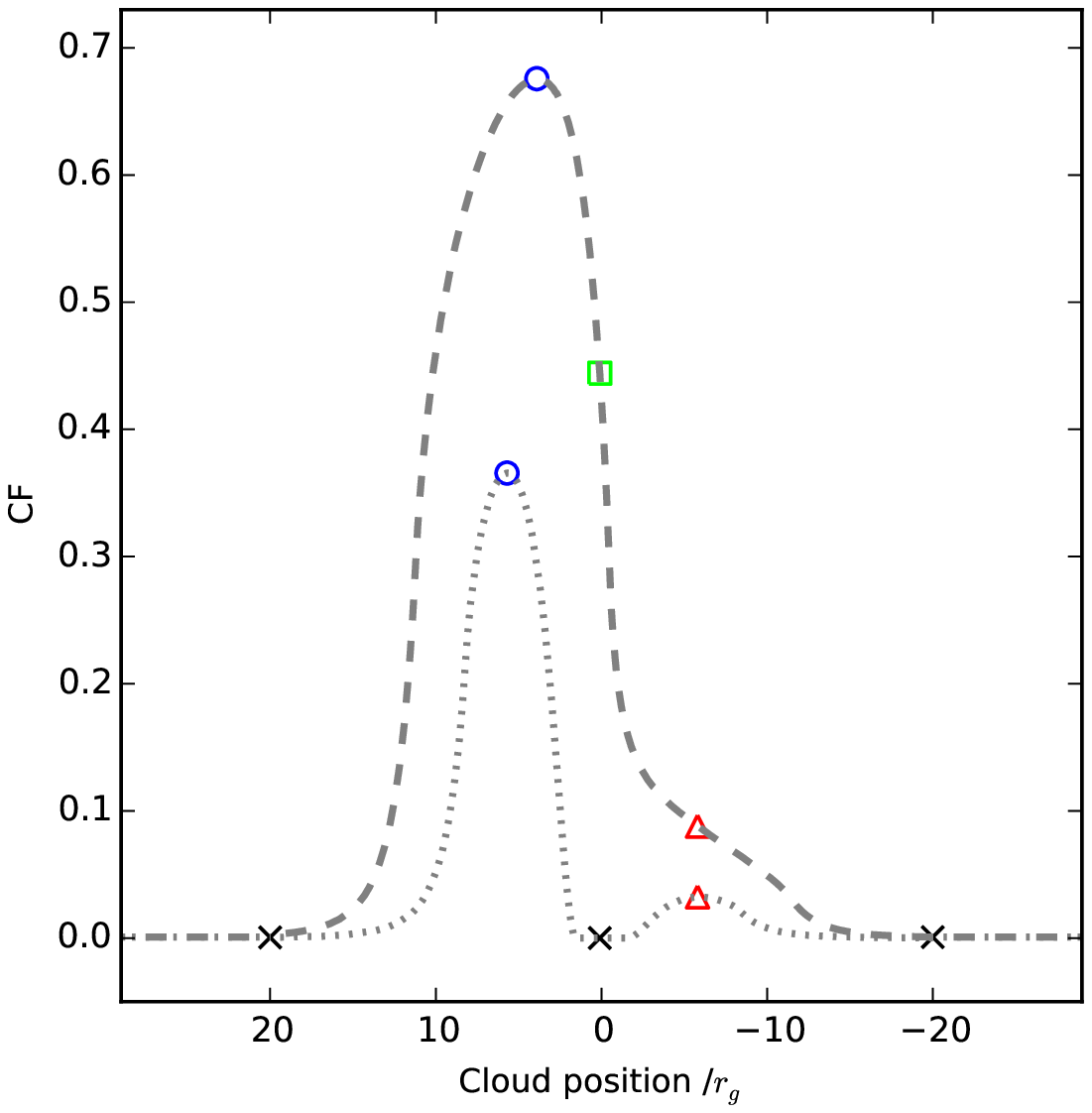}
\includegraphics[height=0.48\textwidth,angle=-90]{AN_rc3.eps}
\includegraphics[height=0.48\textwidth,angle=-90]{AN_rc6.eps}
\caption{\label{fig:cf_with_spectra}
Top: CF--profiles produced by a neutral cloud with $N_{\rm H} = 10^{24}$\,cm$^{-2}$. The system's inclination is $40 \degr$. The disc reflection has an emissivity index of ${\rm q} = 6$. The black hole spin is ${\rm a} = 0.5$, producing an ISCO of $4.23\,r_{\rm g}$. The cloud size is $3\,r_{\rm g}$ for the dotted line and $6\,r_{\rm g}$ for the dashed line. 
Middle: Simulated spectra during a $3\,r_{\rm g}$--sized cloud eclipse. 
Bottom: Simulated spectra during a $6\,r_{\rm g}$--sized cloud eclipse. 
The shapes\,/\,colours code is the same for the three panels, and indicates the location of the cloud (top panel) producing each one of the spectra (middle and bottom panels). 
The data are rebinned for clarity reasons only. 
See Section\,\ref{subsec:simulated_spectra} for a detailed explanation. }
\end{center}
\end{figure}

Next we generate the corresponding CF--profiles for the situation under study, which are shown in the upper panel of Fig.\,\ref{fig:cf_with_spectra}. 
Beyond the covering fraction profiles, it is more interesting to determine whether a cloud involved in this kind of eclipses could imprint any effect on the spectra taken all along one obscuration event, and if so, discern if it would be detectable by current or future observatories. To this extent, we perform several 10--ks simulations of spectra as they would be observed with the EPIC-pn instrument on board \xmm\ during different stages of an eclipse caused by a BLR cloud of $3\,r_{\rm g}$ and $6\,r_{\rm g}$. We show the most characteristic ones in the middle and bottom panels of Fig.\,\ref{fig:cf_with_spectra}. 

In both cases, the spectra out of the eclipse (at $20\,r_{\rm g}$) are shown in black X--shaped markers. The maximum absorption occurs when the cloud covers the approaching part of the inner accretion disc, at $5.7\,r_{\rm g}$ in the case of the smallest cloud and at $3.9\,r_{\rm g}$ in the case of the biggest, shown in blue circles. The hard band ($2-10$\,\kev) remains unabsorbed during all the event, while the soft band ($0.7-2$\,\kev) shows variability. Since the radius of the small cloud is lower than the ISCO, when the cloud is around the $0\,r_{\rm g}$ position in the reference grid with respect to the black hole, it does not cover any flux, hence the spectrum is again unabsorbed (not shown for clarity, but consistent with the black spectrum in the middle panel of Fig.\,\ref{fig:cf_with_spectra}). However, when the cloud is larger than the ISCO, it intercepts flux from both the approaching and the receding parts of the disc, and absorption is still notable (though lower than when the cloud was in the $3.9\,r_{\rm g}$ position), as shown in green squares in the bottom panel of Fig.\,\ref{fig:cf_with_spectra}. When the cloud is over the receding part of the disc ($-5.9\,r_{\rm g}$), absorption enhances as expected in the case of the small cloud, as shown in red triangles in the middle panel of Fig.\,\ref{fig:cf_with_spectra}, although not as much as when the approaching part was eclipsed. In the case of the biggest cloud, when it occupies the same position absorption does not enhance since the cloud is larger than the ISCO and covers part of the inner accretion disc all the time. 
Finally, as the cloud goes away the spectra recover their initial flux values at all energy ranges (not shown for clarity reasons, but consistent with the black spectra). From this analysis, specifically of the small cloud transit, we conclude that even though the CF is low, both eclipses are deep enough to be detected spectroscopically. 

\subsection{Model--independent H\,/\,S light curves}
\label{subsec:HS}

\begin{figure}
\begin{center}
\includegraphics[width=0.4\textwidth]{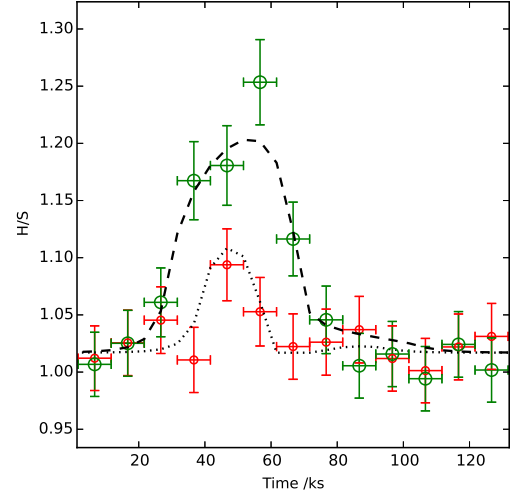}
\caption{\label{fig:hs}
H\,/\,S light curves produced by the same clouds as in Fig.\,\ref{fig:cf_with_spectra}. 
Hard band: 2--10\,\kev . 
Soft band: 0.7--2\,\kev . 
The circles represent the net hardness ratios computed from the count rates of the simulated spectra. The cloud radius is $3\,r_{\rm g}$ for the small red marks, while it is $6\,r_{\rm g}$ for the large green ones. The model--predicted ratios are shown as dotted and dashed lines for the small and large clouds respectively. See Section\,\ref{subsec:HS}. }
\end{center}
\end{figure}

It is even more interesting to compare the different behaviour between the soft and the hard bands: while the clouds are transparent to hard photons, part of the soft ones are absorbed. Therefore these kind of obscuration events drive absorption variability in the soft \xr\ band. With the aim of better understanding how this effect takes place, we simulate the H\,/\,S ratios too. The process is as follows: from the spectra generated in Section\,\ref{subsec:simulated_spectra} we compute the count rates within the hard band in the $2-10$\,\kev\ range, and also in the soft band in the $0.7-2$\,\kev\ range, and then we compute the H\,/\,S ratio light curve of each observation simply by computing the fraction of the hard band over the soft one. 
The light curves are shown in Fig.\,\ref{fig:hs}, in terms of the net count rates computed from the simulated spectra (red small circles for the $3\,r_{\rm g}$--sized cloud, green large circles for the $6\,r_{\rm g}$--sized cloud), together with the model--predicted count rates (dotted and dashed lines respectively). 
The advantage of this technique is that the eclipse can be identified and characterised just computing the total count rates in different bands of each one of the spectra, in a completely model--independent way. By comparing Fig.\,\ref{fig:cf_with_spectra} and Fig.\,\ref{fig:hs}, we conclude that an obscuration event is associated with the existence of high H\,/\,S ratio regions in the light curves. In addition, the H\,/\,S light curves and the CF profiles share the same shape, so that the eclipse parameters could be measured in a model--independent way studying the H\,/\,S ratio of several short spectra all along one observation. 
As an example, from the $3\,r_{\rm g}$ CF--profile in Fig.\,\ref{fig:cf_with_spectra}, we find the distance between the two maxima to be $11.6\pm0.1\,r_{\rm g}$, which corresponds with a time interval of $38.7\pm0.5$\,ks for a cloud moving at 3000\,\kms . From the $3\,r_{\rm g}$ H\,/\,S light curve we find this interval to be $40 \pm 7$\,ks for the same cloud, which is consistent with the value computed from the model--dependent CFs. 
Designing a finer--scale sampling is not possible for the chosen set of parameters, but it would not be an issue to do it for a system with a slower cloud or a larger black hole mass. We let the study of such systems for a more exhaustive future paper (Sanfrutos \etal\ in preparation). 

%
\section{Discussion}
\label{sec:discussion}

We have shown some promising preliminar results. 
In Section\,\ref{sec:absorption_model} we defined a model for the continuum \xr\ emission plus the reflection component convolved with the relativistic kernel {\small KYNCONV} \citep{dovciak2004}, giving account of GR effects due to the proximity of the material in the inner accretion disc to the SMBH. The eclipse of these regions by a Compton--thick neutral cloud produces peculiar CF--profiles depending on the inclination of the system, on the black hole spin and emissivity index, and on the size of the absorber. The study of such profiles allows us to discern how much flux is emitted from every region within the disc. This way we find that the emission measured from our point of view is anisotropic: it is much larger from the parts of the disc that are running towards us in their orbital motion around the SMBH than from its receding parts. This is an effect of Doppler boosting, and its detection would definitely prove that GR effects imprint the AGN \xr\ spectra. We may have already detected these disc emission anisotropies during some of these BLR clouds eclipses, since the shapes of the CF--profiles shown in Fig.\,\ref{fig:cf_profiles} are remarkably similar to those shown in Fig.\,6 in \cite{Risaliti09} and Fig.\,6 in \cite{sanfrutos2013}. This encourages us to deepen the study of these relativistic effects by fitting simulated and real spectra, with the caveat of the detectability limit. 

In Section\,\ref{subsec:simulated_spectra} we explored the detectability of GR effects in a set of simulated \xr\ spectra along one occultation event, finding that (i) a Compton-thick ($N_{\rm H} = 10^{24}$\,cm$^{-2}$) neutral cloud is transparent to hard \xr s ($2-10$\,\kev), but produces absorption in the soft band ($0.7-2$\,\kev); and that (ii) the maximum absorption occurs when the cloud intercepts our LOS to the approaching part of the accretion disc, as expected due to Doppler boosting. 
The difference in absorption between the hard and the soft bands led us to study the H\,/\,S ratio light curves produced during an eclipse. In Section\,\ref{subsec:HS} we showed that the behaviour of the H\,/\,S light curves follows that of the CF--profiles, hence we concluded that it is possible to characterise the properties of the innermost regions of the accretion disc in a new model--independent way just by computing the count rates of the hard and soft bands of spectra taken all along an eclipse. 

Only \xmm\ simulations were studied at the moment. In a forthcoming work (Sanfrutos \etal\ in preparation) we will fit real \xmm\ data and will also simulate spectra from other instruments. In this context we will study the effect of eclipses produced by thicker and\,/\,or ionised clouds in harder bands by means of \nustar. We will also perform simulations of spectroscopic data from \athena, considering more complex models giving account of resolved spectral absorption lines \citep[{\small PHASE} code by][]{krongold_phase2003} in order to take advantage of the outstanding spectral resolution of the future \xr\ observatory. 

Finally, among the technical issues to take into account, it will be important to refine any possible degeneracies among the parameters, to study spectra in several bands of interest in the \xmm\ archive, and to study variations in the soft excess band, where the sensitivity of our instruments gets its highest.

Financial support for this work was provided by the European Union through the Seventh Framework Programme (FP7/2007--2013) under grant n. 312789. 
MS thanks CSIC for support through a JAE--Predoc grant, and rejects cuts to public science budget.

\bibliographystyle{aa}
\bibliography{references}

\end{document}